\documentclass{article}

\usepackage{graphicx}
 \usepackage[preprint]{neurips_2026}

\usepackage[utf8]{inputenc} 
\usepackage[T1]{fontenc}    
\usepackage{hyperref}       
\usepackage{url}            
\usepackage{booktabs}       
\usepackage{amsfonts}       
\usepackage{nicefrac}       
\usepackage{microtype}      
\usepackage{xcolor}         

\title{Towards Foundation Models\\on Hardware Accelerators for Particle Physics}

\newcommand{\authblock}[2]{%
  \parbox[t]{0.45\linewidth}{\centering\textbf{#1}\\ {\normalfont #2}}}

\author{%
\authblock{Maya Benyas}{Stanford University}
\And
\authblock{Julia Gonski}{SLAC National Accelerator Laboratory}
\AND
\authblock{Qibin Liu}{SLAC National Accelerator Laboratory}
\And
\authblock{P. Alex May}{University of California, Irvine}
\AND
\authblock{Vinicius Mikuni}{Nagoya University, Kobayashi-Maskawa Institute}
\And
\authblock{Benjamin Nachman}{Stanford University\\ SLAC National Accelerator Laboratory}
\AND
\authblock{Tanvi Wamorkar}{Stanford University\\ \texttt{twamorka@stanford.edu}}
\AND
\authblock{Liangyu Wu}{Stanford University}
}

\begin{document}

\maketitle

\begin{abstract}
Bandwidth constraints require many particle physics experiments to make real-time decisions on custom hardware or firmware running simplified algorithms, unlike offline analysis, where latency is usually not a limiting factor.
For example, for particle jet tagging at colliders, state-of-the-art performance is achieved by foundation models with hundreds of millions of parameters pre-trained with billions of jets. 
    We use knowledge distillation to transfer what
    such models have learned into efficient networks towards deployment in hardware accelerators. The teacher is the OmniLearned foundation model fine-tuned on
    top quark jet tagging; the student is an attention-free Deep Sets network. We
    demonstrate three ways the student's performance improves: adding a
    message-passing layer to the Deep Sets architecture, training on the
    teacher's soft labels rather than on ground-truth labels alone, and
    distilling from a pretrained teacher rather than from the same
    architecture trained from scratch. In each case the gain is largest in
    the background rejection at low signal efficiency, the regime that
    is most relevant for a trigger.
\end{abstract}

\section{Introduction}
Foundation models pretrained across many jet-physics datasets and tasks now set the accuracy benchmark for jet tagging~\cite{Bhimji:2025isp} and other tasks involving point clouds~\cite{Qu:2022mxj,Qu:2019gqs}. These models have hundreds of millions of parameters, far beyond what a hardware trigger can accommodate. FPGA-based trigger systems at the Large Hadron Collider (LHC) need to make decisions within a few microseconds, requiring machine learning models to be co-designed and compressed for FPGA deployment~\cite{Schulte:2025mai, Sun:2024soe, May:2026cnw,Govorkova_2022}.

Knowledge distillation uses a large teacher's soft predictions to train
a compact student, recovering some of the teacher's accuracy without
its inference cost~\cite{hinton2015distilling}.
Liu et al.~\cite{Liu:2023dio} distilled a Lorentz-equivariant LorentzNet~\cite{Gong:2022lye}
teacher into a 68k-parameter Deep Sets student for top tagging, transferring the
teacher's symmetry bias along with its accuracy, though equivariant architectures
carry trade-offs of their own~\cite{Favaro:2026amj}. \cite{Petitjean:2025zjf} instead builds equivariance directly into
a slim, quantized tagger with no teacher. DistillNet~\cite{Bal:2023bvt} distilled a
graph-neural-network teacher into a compact per-particle network for the CMS Level-1 trigger, reaching about 25~ns FPGA latency
with a small accuracy loss after quantization, and~\cite{Tahseen:2026uaa} distilled a normalizing-flow-based anomaly detector to a similar latency budget. In every case the teacher was trained from scratch on a single task, leaving open what a student gains from a \emph{pretrained} teacher.

In this paper, we distill a 423M-parameter foundation model, fine-tuned on top tagging, into Deep Sets students as small as 17k parameters, and show that a foundation model's accuracy can be transferred to a student based on relatively simple and computationally efficient architectures. Such distillation work lays out the next step of eventual co-design for a hardware accelerator deployment target, with trigger FPGAs being a natural first example.

\section{Method}

  \subsection{Dataset}
  We use the top tagging dataset~\cite{Kasieczka:2019dbj}, in which the signal is
  hadronic top-quark jets and the background is light-quark and gluon jets
  with $p_T \in [550, 650]$~GeV, generated with Pythia8
  \citep{Sjostrand:2014zea} and passed through the Delphes detector
  simulation \citep{deFavereau:2013fsa}. It has 1.2M training, 400k
  validation, and 404k test jets; each jet is given as the four-momenta of
  up to 200 constituents. All numbers below are on the held-out test split.

  \subsection{Teacher}
  The teacher follows the architecture of the OmniLearned foundation model~\cite{Bhimji:2025isp},
  specifically the large model with 28 transformer blocks, 32 attention
  heads, a latent dimension of 1024, and 423M trainable weights. It is
  pretrained on a combined dataset of more than one billion jets spanning
  210 classes, then fine-tuned once on top tagging (94.44\% accuracy, AUC
  0.9880 on the test split). Pretraining and fine-tuning are performed
  once, and the teacher is frozen for every distillation run reported in
  this work. Teacher logits are computed once, offline, on the training
  and validation splits and cached, so the teacher is never re-invoked
  during student training.

  \subsection{Distillation loss}
  Following the standard knowledge-distillation formulation~\cite{hinton2015distilling}, the loss is
  \[
    \mathcal{L} = \alpha \, \mathrm{CE}(s, y)
    + \beta \, T^2 \, D_{\mathrm{KL}}\big(\mathrm{softmax}(t/T) \,\|\, \mathrm{softmax}(s/T)\big),
  \]
  where $s$ is the student's logit vector, $y$ is the ground-truth label,
  and $t$ is the teacher's logit vector. $\alpha$ and $\beta$ weight the
  hard-label and soft-label terms and set how much the student learns from
  the ground truth versus the teacher. $T$ is the temperature: both logit
  vectors are divided by $T$ before the softmax. $\alpha$,
  $\beta$, and $T$ are tunable hyperparameters. We use $\alpha$ = $\beta$ = 0.5 and ${T}$= 4 throughout. The Kullback--Leibler term
  is the knowledge-transfer component of the loss: it drives the student's
  softened output distribution toward the teacher's.

  \subsection{Deep Sets student}
  The student is a plain Deep Sets network~\cite{zaheer2017deep}: a shared MLP $\phi$ embeds each
  particle independently, the embeddings are averaged over the jet's real
  (non-padded) particles, and a second MLP $\rho$ maps the pooled vector to
  the two class logits. There is no interaction between particles, so the
  model is permutation-invariant by construction and contains no attention.
  Each particle is represented by four features relative to the jet axis,
  $(\Delta\eta, \Delta\phi, \log p_T, \log E)$. $\phi$ maps these four inputs to 64 dimensions and then to 32, followed by a 32--64--32 block with a skip connection; $\rho$ applies one more such block to the pooled, particle-averaged output, ending in a linear layer to the two class outputs (top vs.QCD). All layers use GELU activations, and each block includes a normalization layer
 (\texttt{DynamicTanh})~\cite{zhu2025transformers}. The student has 10{,}981 parameters.

  A pure Deep Sets network does not see pairwise quantities such as the
  invariant mass or angular separation of two constituents, since particles
  are combined only by averaging. To improve the expressivity of the
  architecture, we add a message-passing layer between the per-particle
  embedding and the averaging. For every pair of constituents $(i,j)$, a
  small network takes the two particle representations together with three
  pairwise features, $\log m_{ij}$, $\log \Delta R_{ij}$, and
  $\log k_{T,ij}$, and produces a 32-dimensional message. Each particle's
  representation is then updated with the average of its incoming
  messages. The output remains permutation-invariant. Since the number of
  pairs grows as $K^2$, the message-passing layer keeps only the $K{=}64$
  highest-$p_T$ constituents of each jet. $K{=}64$ was not optimized.

  \section{Results}
  \begin{itemize}

  \item Effect of adding a message-passing layer: The two Deep Sets
  students have the same layer sizes, inputs, and distillation procedure;
  the only difference is the message-passing layer inserted between the
  per-particle network and the averaging (Table~\ref{tab:main}). This
  layer raises the student accuracy from 92.85\% to 93.98\%, the AUC from
  0.9800 to 0.9859, and the background rejection from 748 to 1980 at 30\%
  signal efficiency and from 199 to 450 at 50\%. The cost is an increase
  in model size from 11k to 17k parameters. The larger student is within
  0.4 accuracy points of OmniLearned-Small with 155 times fewer parameters.

  \begin{table}[t]
    \caption{Top-tagging test results for the teacher, the reference
    model, and the Deep Sets students.}
    \label{tab:main}
    \centering
    \small
    \begin{tabular}{lcccc}
      \toprule
      Model & Acc & AUC & $1/\mathrm{FPR}_{30\%}$ & $1/\mathrm{FPR}_{50\%}$ \\
      \midrule
      Teacher: OmniLearned-Large, fine-tuned to top tagging   & 94.44\% & 0.9880 & 3365 & 645 \\
      Reference: OmniLearned-Small, fine-tuned to top tagging & 94.38\% & 0.9875 & 2556 & 577 \\
      \midrule
      Deep Sets (no GNN layer), with distillation                & 92.85\% & 0.9800 & 748  & 199 \\
      Deep Sets (including GNN layer), CE only (no distillation) & 93.82\% & 0.9846 & 1216 & 320 \\
      Deep Sets (including GNN layer), with distillation         & 93.98\% & 0.9859 & 1980 & 450 \\
      \bottomrule
    \end{tabular}
  \end{table}

  \begin{figure}[t]
    \centering
    \includegraphics[width=0.65\linewidth]{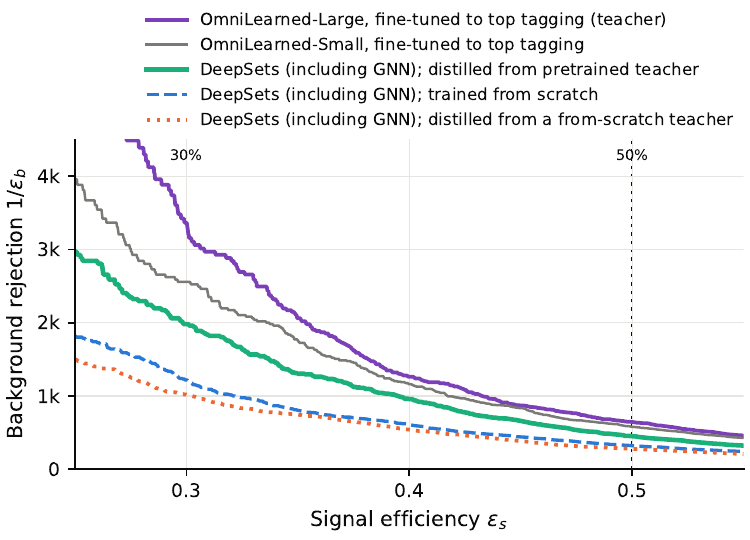}
    \caption{Background rejection versus signal efficiency on the 404k-jet
    test split, zoomed in on the 25–55\% range to show the 30\% and 50\% signal-efficiency operating points.}
    \label{fig:roc}
  \end{figure}

  \item Distillation versus training from scratch: Table~\ref{tab:main}
  compares the same Deep Sets network trained with and without the
  teacher. Trained on hard labels alone, it reaches 93.82\% accuracy with
  a background rejection of 1216 at 30\% signal efficiency and 320 at
  50\%. Distilled from the teacher's soft labels, it reaches 93.98\% with
  rejections of 1980 and 450. Distillation therefore adds 0.16 accuracy
  points and improves the rejection by a factor of 1.6 at 30\% efficiency
  and 1.4 at 50\%. The gain is concentrated at low signal efficiency,
  where the soft labels sharpen the tail of the classifier output
  (Figure~\ref{fig:roc}).

  \begin{table}[t]
    \caption{Effect of teacher pretraining on the distilled student. The
    student is the Deep Sets model with one message-passing layer and the
    teacher is the large OmniLearned model. The same student is distilled
    with the same recipe from two versions of the teacher: the foundation
    model, pretrained and then fine-tuned on top tagging, and a copy of the
    same architecture trained from scratch on top tagging.}
    \label{tab:pretrain}
    \centering
    \small
    \begin{tabular}{lcccc}
      \toprule
      Model & Acc & AUC & $1/\mathrm{FPR}_{30\%}$ & $1/\mathrm{FPR}_{50\%}$ \\
      \midrule
      Student, distilled from the pretrained teacher   & 93.98\% & 0.9859 & 1980 & 450 \\
      Student, distilled from the from-scratch teacher & 93.65\% & 0.9837 & 1020 & 278 \\
      \bottomrule
    \end{tabular}
  \end{table}

  \item Teacher pretraining improves the student's background rejection:
  We distill the same student from two teachers of the same architecture:
  the foundation model, pretrained on the one-billion-jet dataset and then
  fine-tuned on top tagging, and a copy trained from scratch on top tagging
  alone (Table~\ref{tab:pretrain}). The two teachers are close in accuracy
  and AUC but far apart in background rejection: pretraining raises the
  teacher's rejection by a factor of four at 30\% signal efficiency and by
  nearly three at 50\%. The students follow the same pattern. They are
  close in accuracy and AUC but differ in rejection by a factor of two at
  30\% and 1.6 at 50\% signal efficiency. The benefit of pretraining is
  concentrated in the tail of the ROC curve, and distillation passes a
  substantial share of it to the student.
  \end{itemize}

  \subsection{Towards hardware implementation}
  As a step toward studying deployment in hardware accelerators, we quantize both deep sets students to 8 bits and evaluate the performance loss (Table~\ref{tab:quant}). We first apply post-training quantization (PTQ), rounding the weights and inputs of each linear layer to 8 bits without retraining. Each weight matrix and each layer input is rounded to 8-bit integers with one step size per tensor. For the inputs the step size is chosen so that 99.8\% of the values seen on validation jets fit in range; the
  rest are clipped. PTQ reduces the plain student's accuracy by 10 percentage points and the message-passing student's accuracy by 35 points. The latter reaches 59\% accuracy but retains an AUC of 0.92. At 4 bits, both students perform at chance level. We then apply quantization-aware training (QAT), fine-tuning the float32 checkpoints for 15 epochs with 8-bit weights and activations in the forward pass, using Brevitas~\cite{brevitas}. QAT recovers nearly all of the accuracy loss. Relative to float32, the plain student loses 0.27\% points in accuracy and 10\% in background rejection; the message-passing student loses 0.03 points and 4–10\%, respectively. These results indicate that QAT is needed at this model size and that the message-passing layer remains viable under 8-bit quantization. At the median jet multiplicity, the plain student needs 305k multiply-accumulates (MACs) per jet, the Deep Sets + GNN student needs 14.3M, and the teacher needs 37.9G: a three to four order-of-magnitude reduction. MACs measure synthesis cost more directly than parameter count, since FPGAs execute them directly in hardware. The next step is to synthesize the QAT students for FPGA implementation and assess resource and latency estimates; this is work in progress.

\begin{table}[t]
    \caption{Effect of 8-bit quantization on the distilled Deep Sets
    students. MACs are per jet at the median top-tagging multiplicity
    ($N{=}47$) and depend only on the architecture, so they are unchanged
    by quantization; the teacher needs 37.9G MACs per jet for comparison.}
    \label{tab:quant}
    \centering
    \small
    \begin{tabular}{lccccc}
      \toprule
      Model & Acc & AUC & $1/\mathrm{FPR}_{30\%}$ & $1/\mathrm{FPR}_{50\%}$ & MACs \\
      \midrule
      Deep Sets (no GNN layer), float32   & 92.85\% & 0.9800 & 748  & 199 & 305k \\
      Deep Sets (no GNN layer), 8-bit PTQ & 82.90\% & 0.9151 & 45   & 20  & 305k \\
      Deep Sets (no GNN layer), 8-bit QAT & 92.58\% & 0.9785 & 699  & 178 & 305k \\
      \midrule
      Deep Sets (including GNN layer), float32   & 93.98\% & 0.9859 & 1980 & 450 & 14.3M \\
      Deep Sets (including GNN layer), 8-bit PTQ & 59.34\% & 0.9154 & 76   & 29  & 14.3M \\
      Deep Sets (including GNN layer), 8-bit QAT & 93.95\% & 0.9856 & 1787 & 433 & 14.3M \\
      \bottomrule
    \end{tabular}
  \end{table}

 \section{Conclusion}

We distilled a 423M-parameter foundation model into Deep Sets networks with 11k and 17k parameters for top tagging. The addition of one message-passing layer increases background rejection by factors of 2.6 and 2.3 at 30\% and 50\% signal efficiency, respectively. For this architecture, distillation improves rejection by a further factor of 1.6 and 1.4 relative to training on ground-truth labels alone. Distilling from a pretrained teacher instead of the same teacher architecture trained from scratch increases rejection by factors of 1.9 and 1.6. The gains in rejection are substantially larger than the corresponding changes in classification accuracy.

To study the prospects for deployment in hardware accelerators, we quantized both students to 8 bits. Post-training quantization causes large performance losses. With quantization-aware training, the message-passing student reaches 93.95\% accuracy, 0.03 percentage points below float32, with background rejection reduced by 10\% at 30\% signal efficiency and 4\% at 50\%. 
High-level synthesis to register-transfer level for FPGA deployment, including resource and latency optimization, is still in progress.

\section*{Code Availability}
   {\raggedright The code for this work is available at
   \url{https://github.com/stanford-ai4physics/fm-distillation}.\par}
\section*{Acknowledgments}
BN, MB, JG, LW, and QL are supported
by the U.S. Department of Energy (DOE), Office of Science under contract DE-AC02-76SF00515. PAM is supported by the U.S. Department of Energy under contract number DE-SC0024518. TW is supported by the National Science Foundation under Grant No. 2311666. VM is supported by JST EXPERT-J, Japan Grant Number JPMJEX2509. This research
used resources of the National Energy Research Scientific Computing Center, a DOE Office of Science User
Facility supported by the Office of Science of the DOE under Contract No. DE-AC02-05CH11231 using NERSC award HEP-ERCAP0035546 and Contract No. DE-AC02-05CH11231 using NERSC award HEP-ERCAP0037461. PAM was affiliated with San Jose State University during the course of this work.
\newpage

\bibliographystyle{unsrt}

\bibliography{tilman, HEPML, other, EEC_ref, ref} 

\end{document}